\documentclass[aps,pre,groupedaddress]{revtex4}

\usepackage{amsmath}
\usepackage{amssymb}
\usepackage{amsfonts}
\usepackage{physics}
\usepackage{bm}
\usepackage{comment}
\usepackage{graphicx}
\usepackage{color}

\newcommand{\e}{\mathrm{e}}
\newcommand{\I}{\text{I}}

\renewcommand{\S}{\text{S}}

\renewcommand{\Pi}{\varPi}

\begin{document}

\title{Analytical Framework for the Approximate Master Equation}
\author{Yu Takiguchi}
\email{takiguchi.yu.st@gmail.com}
\affiliation{Graduate School of Science, Hokkaido University, 10-8, Kita, Sapporo, 060-0810, Japan}
\author{Takehisa Hasegawa}
\email{takehisa.hasegawa.sci@vc.ibaraki.ac.jp}
\affiliation{Graduate School of Science and Engineering, Ibaraki University, 2-1-1, Bunkyo, Mito, 310-8512, Japan}

\begin{abstract}
The approximate master equation (AME) provides a highly accurate description of dynamical processes on networks, yet its steady states are generally analytically intractable.
In this study, we develop an analytical framework to derive the steady states of the AME by introducing a controlled approximation that enables closure of the moment equations.
This framework reproduces the steady state of the pair approximation by achieving closure with the minimum required order of moments, and can be systematically refined to approach the exact steady states of the AME.
We apply this to the SIS model, the voter model, and evolutionary games, demonstrating that the steady states can be derived.
In particular, for evolutionary games, we show that combining our framework with the singular perturbation method enables the analytical derivation of the time evolution.
\end{abstract}

\maketitle

\section{Introduction}

A wide range of collective phenomena, including rumor propagation, consensus formation, the transmission of infectious diseases, and the diffusion of innovations, have been studied as dynamical processes on networks. 
In such models, nodes represent individuals, and their states evolve through interactions with neighboring nodes.
Due to their inherently many-body nature, exact analytical treatments are generally intractable, making approximate methods indispensable.
The mean-field approximation reduces the dynamics to a few ordinary differential equations, but neglects correlations between neighboring nodes, leading to qualitatively incorrect predictions when such correlations play a crucial role~\cite{Szabo2005,Takiguchi2024}. 
To overcome this limitation, the approximate master equation (AME)~\cite{Gleeson2011,Gleeson2013,Fennell2019} accounts for not only the state transitions of focal nodes but also those of their neighbors, thereby providing a more accurate description of dynamics on networks. 
The AME has been shown to outperform other extensions of the mean-field approach, such as heterogeneous mean-field and pair approximations~\cite{Gleeson2013}. 
It has also been successfully applied to a variety of dynamical processes, including epidemic spreading processes such as the SIS model~\cite{Lindquist2011,Hasegawa2016,Hasegawa2018}, opinion dynamics~\cite{Gleeson2013}, information diffusion models such as the Watts threshold model~\cite{Watts2002,Ruan2015,Kobayashi2023}, and evolutionary games on networks~\cite{Kobayashi2023,Wang2023,Takiguchi2024}.

The high accuracy of the AME comes at the expense of analytical tractability, and thus it has been used primarily as a numerical tool.
Only a limited number of studies have obtained analytical results. 
For example, Lindquist et al.~\cite{Lindquist2011} calculated the critical point for the SIR model using equations equivalent to the AME, and Takiguchi and Nemoto~\cite{Takiguchi2024} derived the critical points for public goods games on networks.
Although critical points can often be obtained from the stability of trivial steady states, no general analytical method is currently available to determine non-trivial steady states of the AME, making this a fundamental open problem.

In this study, we develop an analytical framework to derive the steady states for dynamical processes on networks by applying a specific approximation to the AME. 
This approach is based on imposing the approximation related to a balance condition at steady state, which is not generally satisfied by the exact steady state of the AME.
Nevertheless, our framework reproduces the steady states of the pair approximation and can be systematically refined to approach the exact steady states of the AME.

The remainder of the paper is organized as follows. 
Section~\ref{sec:derivation} briefly reviews the AME, and Section~\ref{sec:method} introduces our analytical framework.
In Section~\ref{sec:examples}, we apply the proposed framework to the SIS model, the voter model, and evolutionary games to derive their steady states.
For evolutionary games, we show that combining our approach with the singular perturbation method enables us to analytically derive the time evolution.

\section{Brief Review of the Approximate Master Equation}
\label{sec:derivation}

We briefly review the formulation of the approximate master equation (AME) for two-state models, following Gleeson~\cite{Gleeson2013}.
We denote the two possible states of each node by $\S$ and $\I$, corresponding to the susceptible and infected states in the SIS model.
The present study focuses on regular networks, where every node has the same degree $k$, although the AME can also be formulated for networks with arbitrary degree distributions; see Refs.~\cite{Gleeson2013} and~\cite{Fennell2019}. 

We define an $s_m$-node as a node in state $s \in \{\S, \I\}$ that has $m$ neighbors in state $\I$ 
(and thus $k-m$ neighbors in state $\S$), where $m = 0, 1, \dots, k$. 
Let $\rho^s_m$ denote the fraction of $s_m$-nodes.
Then, the total fraction of nodes in state $s$ is given by $\rho^s = \sum_{m=0}^k \rho^s_m$, and the normalization condition is $\sum_{s \in \qty{\S, \I}} \sum_{m=0}^{k} \rho^s_m = 1$.

We denote the transition rate at which an $s_m$-node changes into an $s'_{m'}$-node by $W(s_m \to s'_{m'})$. 
The fraction $\rho^s_m$ changes due to state transitions of both the focal node and its neighbors. 
The time evolution of $\rho^{\S}_m$ is described by the following equation:
\begin{align}
    \dv{t} \rho^{\S}_m
    =
    &- W(\S_m \to \I_m) \rho^{\S}_m + W(\I_m \to \S_m) \rho^{\I}_m \nonumber\\
    &- W(\S_m \to \S_{m+1}) \rho^{\S}_m + W(\S_{m-1} \to \S_m) \rho^{\S}_{m-1} \nonumber\\
    &- W(\S_m \to \S_{m-1}) \rho^{\S}_m + W(\S_{m+1} \to \S_m) \rho^{\S}_{m+1}. \label{eq:master}
\end{align}
The first two terms on the right-hand side represent the state changes of the focal node itself, the next two terms account for neighbor transitions from state $\S$ to $\I$, and the final two terms account for those from state $\I$ to $\S$.
Since we consider state transitions over an infinitesimal time interval, simultaneous transitions of multiple nodes, such as $W (\S_m \to \I_{m+1})$ or $W (\S_m \to \S_{m-2})$, can be neglected.
The transition rate of the focal node, $W(s_m \to s'_m)$, is model-dependent.
For example, in the SIS model with infection rate $\beta$ and recovery rate $\gamma$, $W(\S_m \to \I_m) = \beta m$ and $W(\I_m \to \S_m) = \gamma$.
In the voter model, $W(\S_m \to \I_m) = m / k$ and $W(\I_m \to \S_m) = (k-m) / k$.

In the AME framework, the neighbor transition rates, $W(s_m \to s_{m'})$, are expressed in terms of the focal node transition rates $W(s_m \to s'_m)$.
For $s \in \qty{\S, \I}$, let $\beta^{s}$ be the rate at which an $\S$-node adjacent to an $s$-node becomes $\I$, 
and let $\gamma^{s}$ be the rate at which an $\I$-node adjacent to an $s$-node becomes $\S$. 
Noting that the probability of an $s$-node adjacent to an $\S$-node being in state $s_m$ is $(k-m)\rho^s_m / \sum_{m'=0}^k (k-m')\rho^s_{m'}$, 
the rates $\beta^\S$ and $\gamma^\S$ are defined as follows:
\begin{align}
    \beta^\S 
    &\equiv
    \sum_{m=0}^k \frac{ (k-m) \rho^{\S}_m }{\sum_{m'=0}^k (k-m') \rho^{\S}_{m'}} W(\S_m \to \I_m),\\
    \gamma^\S
    &\equiv
    \sum_{m=0}^k \frac{ (k-m) \rho^{\I}_m }{\sum_{m'=0}^k (k-m') \rho^{\I}_{m'}} W(\I_m \to \S_m).
\end{align}
Since an $\S_m$-node has $k-m$ neighbors in state $\S$ and $m$ neighbors in state $\I$, 
the neighbor transition rates can be approximated as $W(\S_m \to \S_{m+1}) = \beta^\S (k-m)$ and $W(\S_m \to \S_{m-1}) = \gamma^\S m$. 
Combining these, Eq.~\eqref{eq:master} can be written as
\begin{align}
    \dv{t} \rho^{\S}_m
    =
    &- W(\S_m \to \I_m) \rho^{\S}_m + W(\I_m \to \S_m) \rho^{\I}_m \nonumber\\
    &- \beta^\S (k-m) \rho^{\S}_m + \beta^\S (k-m+1) \rho^{\S}_{m-1} \nonumber\\
    &- \gamma^\S m  \rho^{\S}_m + \gamma^\S (m+1) \rho^{\S}_{m+1},
    \label{eq:AME_S}
\end{align}
where we adopt the convention $\rho^{\S}_{k+1} = \rho^{\S}_{-1}=0$. 

Similarly, the time evolution of $\rho^{\I}_m$ is given by:
\begin{align}
    \dv{t} \rho^{\I}_m
    =
    &- W(\I_m \to \S_m) \rho^{\I}_m + W(\S_m \to \I_m) \rho^{\S}_m \nonumber\\
    &- \beta^\I (k-m) \rho^{\I}_m + \beta^\I (k-m+1) \rho^{\I}_{m-1} \nonumber\\
    &- \gamma^\I m  \rho^{\I}_m + \gamma^\I (m+1) \rho^{\I}_{m+1},
    \label{eq:AME_I}
\end{align}
where
\begin{align}
    \beta^\I
    &\equiv
    \sum_{m=0}^k \frac{ m \rho^{\S}_m }{\sum_{m'=0}^k m' \rho^{\S}_{m'}} W(\S_m \to \I_m),\\
    \gamma^\I
    &\equiv
    \sum_{m=0}^k \frac{ m \rho^{\I}_m }{\sum_{m'=0}^k m' \rho^{\I}_{m'}} W(\I_m \to \S_m).
\end{align}
Note that the probability of an $s$-node adjacent to an $\I$-node being in state $s_m$ is $m\rho^s_m / \sum_{m'=0}^k m'\rho^s_{m'}$. 

By numerically solving the above equations, we obtain the fractions of $\S$- and $\I$-nodes at time $t$, denoted by $\rho^{\S}(t)$ and $\rho^{\I}(t)$.
Suppose that the initial fraction of $\I$-nodes is $\rho^{\I}(0) = \rho^{\I}_0$ (and thus $\rho^{\S}(0) = 1 - \rho^{\I}_0$), and that these nodes are distributed uniformly at random over the network.
Defining the binomial distribution as
\begin{align}
    B_m (q) 
    \equiv \binom{k}{m} q^m (1-q)^{k-m},
    \label{eq:binomial}
\end{align}
the initial conditions for the AME can be set as $\rho^{\I}_m(0) = \rho^{\I}_0 B_m (\rho^{\I}_0)$ and $\rho^{\S}_m(0) = (1-\rho^{\I}_0) B_m (\rho^{\I}_0)$.

\subsubsection*{Reduction of the AME to Mean-Field and Pair Approximations}

We discuss how the AME reduces to the mean-field and pair approximations~\cite{Gleeson2013}.
Parenthetically, the reduction remains valid for networks with arbitrary degree distributions.
While the AME describes the time evolution of the fraction of $s_m$-nodes, $\rho^s_m$, the mean-field approximation focuses on the total fraction of $s$-nodes, $\rho^s$.
The mean-field approximation assumes that nodes in states $\S$ and $\I$ are distributed uniformly at random over the network, thereby neglecting correlations between neighboring nodes. 
Substituting $\rho^{\I}_m = \rho^{\I} B_m(\rho^{\I})$ and $\rho^{\S}_m = \rho^{\S} B_m(\rho^{\I})$ in Eq.~\eqref{eq:AME_S}, the AME reduces to the mean-field approximation:
\begin{align}
    \dv{t} \rho^{\S} 
    &= - \sum_{m=0}^k W(\S_m \to \I_m) B_m(\rho^{\I}) \rho^{\S} + \sum_{m=0}^k W(\I_m \to \S_m) B_m(\rho^{\I}) \rho^{\I} \\
    &\equiv - W(\S \to \I) \rho^{\S} + W(\I \to \S) \rho^{\I}. \label{eq:reductionToMeanField}
\end{align}
For the SIS model, for example, the transition rates in Eq.~\eqref{eq:reductionToMeanField} are $W(\S \to \I) = \beta \rho^{\I}$ and $W(\I \to \S) = \gamma$.

The pair approximation describes the time evolution of the fraction of adjacent node pairs in states $s$ and $s'$, denoted by $\rho^{ss'}$.
The total fraction of nodes in state $s$ is given by $\rho^s = \sum_{s'} \rho^{ss'}$, and the conditional probability that a neighbor of an $s'$-node is in state $s$ is defined as $\rho^{s|s'} \equiv \rho^{ss'}/\rho^{s'}$.
While the pair approximation accounts for correlations between adjacent nodes, it neglects correlations between nodes separated by a distance of two or more. 
Substituting $\rho^{\I}_m = \rho^{\I} B_m(\rho^{\I|\I})$ and $\rho^{\S}_m = \rho^{\S} B_m(\rho^{\I|\S})$ in Eqs.~\eqref{eq:AME_S} and \eqref{eq:AME_I}, the AME reduces to the pair approximation:
\begin{align}
    \dv{t} \rho^{\I\S}
    =
    &- \sum_{m=0}^{k} \frac{2m-k}{k} W(\S_m \to \I_m) B_m(\rho^{\I|\S}) \rho^{\S} \nonumber\\
    &+ \sum_{m=0}^{k} \frac{2m-k}{k} W(\I_m \to \S_m) B_m(\rho^{\I|\I}) \rho^{\I},\\
    \dv{t} \rho^{\I\I}
    = 
    &- \sum_{m=0}^{k} \frac{2m}{k} W(\I_m \to \S_m) B_m(\rho^{\I|\I}) \rho^{\I} \nonumber\\
    &+ \sum_{m=0}^{k} \frac{2m}{k} W(\S_m \to \I_m) B_m(\rho^{\I|\S}) \rho^{\S}.
\end{align}
Note that $\rho^{\I s} = \sum_{m=0}^k m \rho^{s}_m / k$ and $\rho^{\S s} = \sum_{m=0}^k (k-m) \rho^{s}_m / k$. 
The symmetry $\rho^{\I\S} = \rho^{\S\I}$ and the normalization $\rho^{\S\S}+\rho^{\S\I}+\rho^{\I\S}+\rho^{\I\I} = 1$ are satisfied.

\section{Analytical Framework for Deriving Steady States of the Approximate Master Equation}
\label{sec:method}

In the following, we consider a class of models where the transition rates are linear functions of $m$:
\begin{align}
    W(\S_m \to \I_m) &= \sum_{l = 0}^{1} C_{\S,l} m^l,\\
    W(\I_m \to \S_m) &= \sum_{l = 0}^{1} C_{\I,l} m^l,
\end{align}
where $C_{s,l}$ are model-dependent constants. 
For example, the SIS model is characterized by $C_{\S,0} = 0$, $C_{\S,1} = \beta$, $C_{\I,0} = \gamma$, and $C_{\I,1} = 0$. 
The voter model is characterized by $C_{\S,0} = 0$, $C_{\S,1} = 1/k$, $C_{\I,0} = 1$, and $C_{\I,1} = -1/k$.

The steady state of the AME can be obtained by setting $\dv{t} \rho^s_m = 0$ for all $s \in \{\S, \I\}$ and $m = 0, \ldots, k$; these equations are, however, analytically intractable.
To address this, we propose an analytical framework to obtain the steady-state fractions $\rho^s$.
We start by transforming the variables from the fractions $\rho^s_m$ to the moments $\ev*{m^j}_s$, where $\ev*{f(m)}_s \equiv \sum_m f(m)\rho_m^s$.
Note that $\ev*{1}_s = \rho^{s}$, which is not necessarily equal to unity.
We refer to $\ev*{m^j}_s$ as the $j$-th order $s$-moment. 
The fraction of $s$-nodes can be expressed using the zeroth-order $s$-moment as $\rho^{s} = \ev*{m^0}_{s}$, 
and the fractions of node pairs are given by $\rho^{\S\S} = \ev*{k-m}_{\S}/k$, $\rho^{\I\I} = \ev*{m}_{\I}/k$, $\rho^{\S\I} = \rho^{\I\S} = \ev*{m}_{\S}/k = \ev*{k-m}_{\I}/k$.
The fraction of triplets can also be represented in terms of moments.

The time evolution of the $j$-th order $\S$-moments can be derived from Eq.~\eqref{eq:AME_S} as follows:
\begin{align}
    \dv{t} \ev*{m^j}_{\S}
    = 
    &- \sum_{m=0}^k m^j W(\S_m \to \I_m) \rho^{\S}_m + \sum_{m=0}^k m^j W(\I_m \to \S_m) \rho^{\I}_m \nonumber\\
    &- \beta^\S \sum_{m=0}^k m^j (k-m) \rho^{\S}_m + \beta^\S \sum_{m'=0}^{k-1} (m'+1)^j (k-m') \rho^{\S}_{m'} \nonumber\\
    &- \gamma^\S \sum_{m=0}^k m^{j+1}  \rho^{\S}_m + \gamma^\S \sum_{m'=1}^k (m'-1)^j m' \rho^{\S}_{m'} \nonumber\\
    = 
    &- \ev*{ m^j W(\S_m \to \I_m) }_{\S} + \ev*{ m^j W(\I_m \to \S_m) }_{\I} \nonumber\\
    &+ \beta^\S \ev*{ \qty[ (m+1)^j - m^j ] (k-m) }_{\S} - \gamma^\S \ev*{ \qty[ m^j - (m-1)^j ] m }_{\S},
    \label{eq:moment_jS}
\end{align}
Since
\begin{align}
    \ev*{ m^j W(\S_m \to \I_m) }_{\S} &= \sum_{l=0}^1 C_{\S,l} \ev*{m^{j+l}}_{\S},
\end{align}
the first and second terms on the right-hand side of Eq.~\eqref{eq:moment_jS} can be expressed in terms of moments up to order $j+1$. 
The third and fourth terms involve moments up to the second order when $j \leq 1$, 
and up to order $j$ when $j \geq 2$.
Similarly, the time evolution of the $j$-th order $\I$-moments is
\begin{align}
    \dv{t} \ev*{m^j}_{\I}
    =
    &- \ev*{ m^j W(\I_m \to \S_m) }_{\I} + \ev*{ m^j W(\S_m \to \I_m) }_{\S}\nonumber\\
    &+ \beta^\I \ev*{ \qty[ (m+1)^j - m^j ] (k-m) }_{\I} - \gamma^\I \ev*{ \qty[ m^j - (m-1)^j ] m }_{\I}.
    \label{eq:moment_jI}
\end{align}
The first and second terms on the right-hand side can be expressed in terms of moments up to order $j+1$. 
The third and fourth terms involve moments up to the second order when $j \leq 1$, and up to order $j$ when $j \geq 2$.

We determine the moments in the steady state. 
Let $n$ be an arbitrarily chosen natural number such that $n \geq 2$. 
By setting $\dv{t} \ev*{m^j}_{\S} = \dv{t} \ev*{m^j}_{\I} = 0$ for $j=0, 1, \ldots, n$, we obtain a total of $2(n+1)$ independent equations. 
Since these equations involve $2(n+2)$ unknown moments up to order $n+1$, they are not closed for any choice of $n$. 
Increasing $n$ does not resolve this issue, as each additional two equations for a higher-order moment introduces two additional unknowns.

To obtain a closed set of equations for moments up to order $n$, we introduce the following approximation in the steady state for a given $n$:
\begin{align}
    - \ev*{ m^n W(\I_m \to \S_m) }_{\I} + \ev*{ m^n W(\S_m \to \I_m) }_{\S} = 0.
    \label{eq:approximation}
\end{align}
Substituting this relation into $\dv{t} \ev*{m^n}_s = 0$ eliminates the $(n+1)$-th order moments from the equations, thereby closing the system at order $n$.
Then, the $2(n+1)$ equations can be solved to determine all moments up to order $n$. 
Furthermore, these results can be used to estimate higher-order moments.

\subsubsection*{Interpretation as the Semi-Detailed Balance Condition}

In our framework, we assume that the relation~\eqref{eq:approximation} holds in the steady state for a given $n$. 
In contrast, under the pair approximation, the same relation holds for all $j=0, 1, \ldots$ (see Appendix~\ref{sec:pair_approximation}):
\begin{align}
    - \ev*{ m^j W(\I_m \to \S_m) }_{\I} + \ev*{ m^j W(\S_m \to \I_m) }_{\S} = 0.
    \label{eq:approximation_pair}
\end{align}
This is equivalent to a local balance condition, which we call the ``semi-detailed balance condition'':
\begin{align}
    - W(\I_m \to \S_m) \rho^{\I}_m + W(\S_m \to \I_m) \rho^{\S}_m = 0,
    \label{eq:balance_condition}
\end{align}
which holds for all $m = 0, 1, \ldots, k$ (see Appendix~\ref{sec:balance_condition}).
In the steady state of the AME, the total inflow and outflow of $\rho^s_m$ must balance, including contributions from state transitions of neighboring nodes. 
However, Eq.~\eqref{eq:balance_condition} represents a balance arising solely from the state transitions of the focal node itself. 
The semi-detailed balance condition is more restrictive than the global balance condition but less so than the detailed balance condition.
Generally, dynamical processes on networks do not satisfy the semi-detailed balance condition in the steady state. 
Since the pair approximation restricts the steady state to those satisfying this condition, it introduces a systematic bias and thus generally deviates from the true steady state.
In our approach, we assume Eq.~\eqref{eq:approximation_pair} only for $j=n$. 
By relaxing the constraint in this way, our method enables higher accuracy than the pair approximation.

From Eq.~\eqref{eq:moment_jI}, we obtain:
\begin{align}
    \dv{t} \ev*{m^0}_{\I}
    =
    &- \ev*{ W(\I_m \to \S_m) }_{\I} + \ev*{ W(\S_m \to \I_m) }_{\S},\\
    \dv{t} \ev*{m}_{\I}
    =
    &- 2\ev*{ m W(\I_m \to \S_m) }_{\I} + 2\ev*{ m W(\S_m \to \I_m) }_{\S}.
\end{align}
This implies that Eq.~\eqref{eq:approximation_pair} holds automatically for $j=0,1$ in the steady state of the AME. 
Setting $n=2$ and employing Eq.~\eqref{eq:approximation}, we obtain a closed set of equations for moments up to second order. 
Since the formulation involves only equations that are also satisfied under the pair approximation, the resulting steady state coincides with that obtained from the pair approximation.
For $n \geq 3$, however, the results deviate from those of the pair approximation, since Eq.~\eqref{eq:approximation_pair} does not hold for $j=2, \ldots, n-1$. 
The recovery of the pair approximation using moments up to second order indicates that this approximation only accounts for the mean and variance of $m$, the number of neighbors in state $\I$. 
Setting a larger $n$ allows the model to capture the influence of higher-order moments, such as skewness and kurtosis.

\section{Examples}
\label{sec:examples}

To demonstrate the validity and versatility of the proposed framework, we now apply it to several representative models, including the SIS model, the voter model, and evolutionary games.

\subsection{SIS Model}
\label{sec:SIS_model}

\begin{figure}[htbp]
    \centering
    \includegraphics[width=0.8\linewidth]{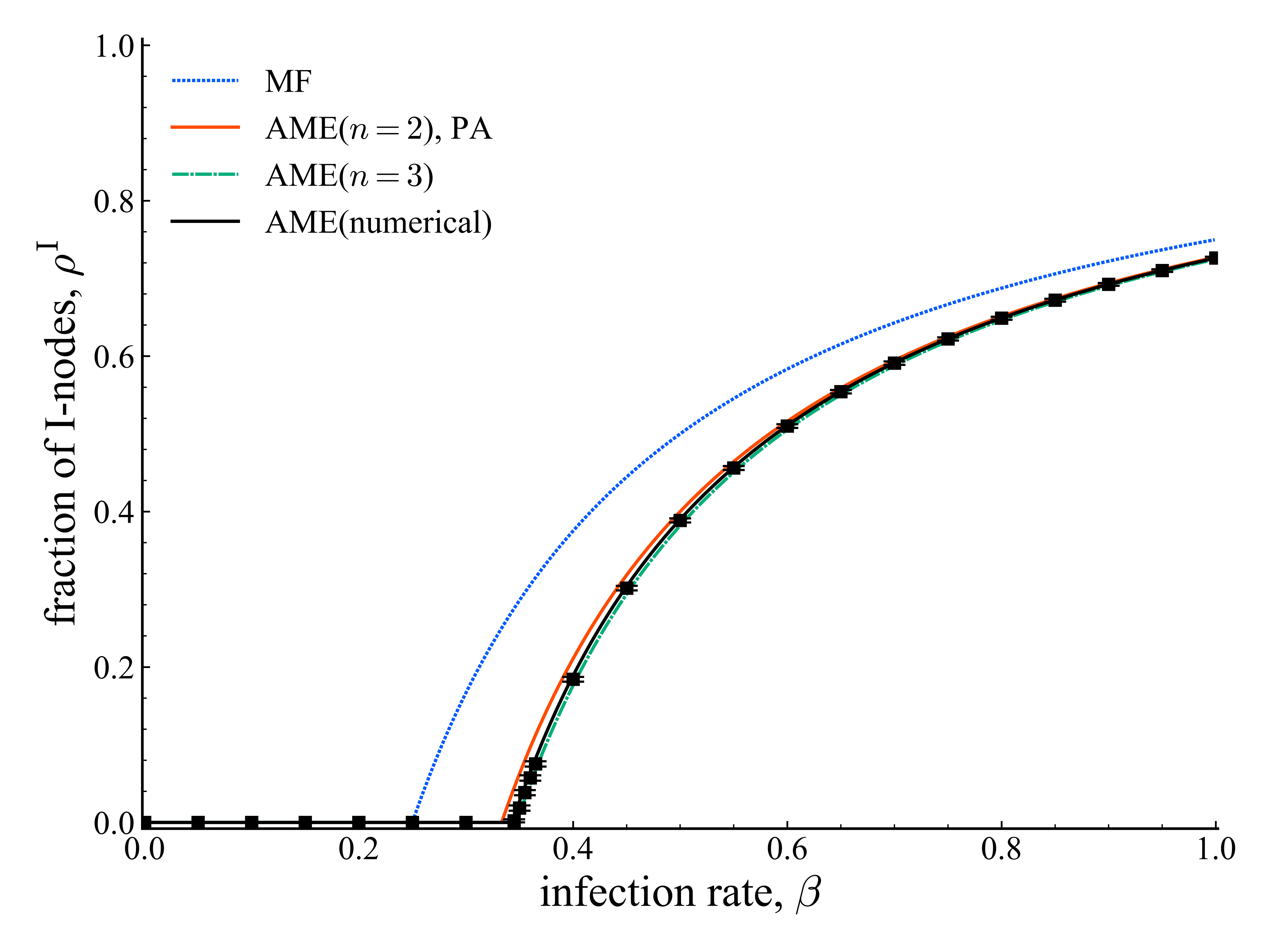}
    \caption{
        Steady state of the SIS model with degree $k=4$ and recovery rate $\gamma=1$. 
        MF, PA, and AME $(n=2,3)$ denote the mean-field approximation, pair approximation, and the AME with moments up to order $n$, respectively.
	    The black solid line represents the numerical solution of the AME.
        Black squares represent Monte Carlo simulation results on a regular random graph with $10^5$ nodes. 
        Monte Carlo results are averaged over $10^3$ independent runs, with error bars indicating the standard deviation.
        \label{fig:SIS_model}
    }
\end{figure}

We first consider the SIS model, a fundamental framework for describing the spread of infectious diseases.
Each node can be in one of two states: susceptible ($\S$) or infected ($\I$).
$\S$-nodes can be infected, whereas $\I$-nodes can transmit the disease to their neighbors.
Let $\beta$ and $\gamma$ denote the infection and recovery rates, respectively.
During an infinitesimal time interval $\dd t$, an $\S$-node is infected by each of its $\I$-neighbors independently with probability $\beta \dd t$, while an $\I$-node recovers to state $\S$ with a constant probability $\gamma \dd t$.
The transition rates for a focal node are given by $W(\S_m \to \I_m) = \beta m$ and $W(\I_m \to \S_m) = \gamma$.
From Eqs.~\eqref{eq:moment_jS} and~\eqref{eq:moment_jI}, evaluated for $j=0, 1, 2$, the following six equations hold in the steady state:
\begin{align}
    1 &= \ev*{m^0}_{\S} + \ev*{m^0}_{\I},
    \label{eq:moment_SIS_1}\\
    0 &= - \beta \ev*{m}_{\S} + \gamma \ev*{m^0}_{\I},
    \label{eq:moment_SIS_2}\\
    0 &= \ev*{m}_{\S} + \ev*{m}_{\I} - k \ev*{m^0}_{\I},
    \label{eq:moment_SIS_3}\\
    0 &= - \beta \ev*{m^2}_{\S} + \gamma \ev*{m}_{\I},
    \label{eq:moment_SIS_4}\\
    0 &=
    - \beta \ev*{m^3}_{\S} + \gamma \ev*{m^2}_{\I} \nonumber\\
    &\qquad
    + \beta \frac{ k \ev*{m}_{\S} - \ev*{m^2}_{\S} }{ k \ev*{m^0}_{\S} - \ev*{m}_{\S} } \qty[ - 2 \ev*{m^2}_{\S} + (2k-1) \ev*{m}_{\S} + k \ev*{m^0}_{\S} ]
    - 2 \gamma \ev*{m^2}_{\S} + \gamma \ev*{m}_{\S},
    \label{eq:moment_SIS_5}\\
    0 &= 
    + \beta \ev*{m^3}_{\S} - \gamma \ev*{m^2}_{\I} \nonumber\\
    &\qquad
    + \beta \frac{\ev*{m^2}_{\S}}{\ev*{m}_{\S}} \qty[ - 2 \ev*{m^2}_{\I} + (2k-1) \ev*{m}_{\I} + k \ev*{m^0}_{\I} ]
    - 2 \gamma \ev*{m^2}_{\I} + \gamma \ev*{m}_{\I}. 
    \label{eq:moment_SIS_6}
\end{align}
Here, Eq.~\eqref{eq:moment_SIS_1} represents the normalization condition $\rho^{\S} + \rho^{\I} = 1$, Eq.~\eqref{eq:moment_SIS_2} corresponds to the global balance condition, and Eq.~\eqref{eq:moment_SIS_3} expresses the symmetry $\rho^{\S\I} = \rho^{\I\S}$. 
Note that the above set of equations is not closed because it involves third-order moments. 
By using Eq.~\eqref{eq:approximation} with $n=2$:
\begin{align}
    0 &= - \beta \ev*{m^3}_{\S} + \gamma \ev*{m^2}_{\I},
    \label{eq:moment_SIS_7}
\end{align}
we obtain the stationary fraction of infected nodes, $\rho^{\I} = \ev*{m^0}_{\I}$, as follows:
\begin{align}
    \rho^{\I} &= 0,\ \frac{ ( k - 1) \frac{\beta}{\gamma} - 1 }{ (k - 1) \frac{\beta}{\gamma} - \frac{1}{k} }.
\end{align}
This result coincides with that obtained via the pair approximation~\cite{Luo2014}.

We next consider moments up to the third order.
Using Eq.~\eqref{eq:approximation} with $n=3$, we obtain the following steady state:
\begin{align}
    \rho^{\I} &= \frac{a_1 - \sqrt{ a_1^2 - 4 a_2 a_0 }}{2 a_2},\\
    a_2 &\equiv - 2 \frac{\gamma^3}{\beta^3} + k ( 2k - 5 ) \frac{\gamma^2}{\beta^2} + 2 k^2 ( k-1 ) \frac{\gamma}{\beta}  - k^3 ( k - 1 ),\\
    a_1 &\equiv - 2 k \frac{\gamma^3}{\beta^3} - k (6-k) \frac{\gamma^2}{\beta^2} + 6 k^2 (k-1) \frac{\gamma}{\beta} - 2 k^3 (k-1),\\
    a_0 &\equiv - k \frac{\gamma^3}{\beta^3} - 2 k^2 \frac{\gamma^2}{\beta^2} + 4 k^2 (k-1) \frac{\gamma}{\beta} - k^3 (k-1). 
\end{align}

Figure~\ref{fig:SIS_model} shows the steady states obtained by each approximation, along with Monte Carlo simulation results.
The mean-field approximation (blue dotted line) exhibits a significant deviation from the numerical solution of the AME (black solid line).
The result for $n=2$ (red solid line), which coincides with the pair approximation, improves the accuracy, although small deviations remain.
In contrast, the result for $n=3$ (green dash-dotted line) is in good agreement with both the numerical solution of the AME and Monte Carlo simulation results (black squares). 
Furthermore, numerical evaluation of the $n=4$ approximation yields results even closer to the exact steady state of the AME.

\subsection{Voter Model}
\label{sec:voter_model}

\begin{figure}[htbp]
    \centering
    \includegraphics[width=0.8\linewidth]{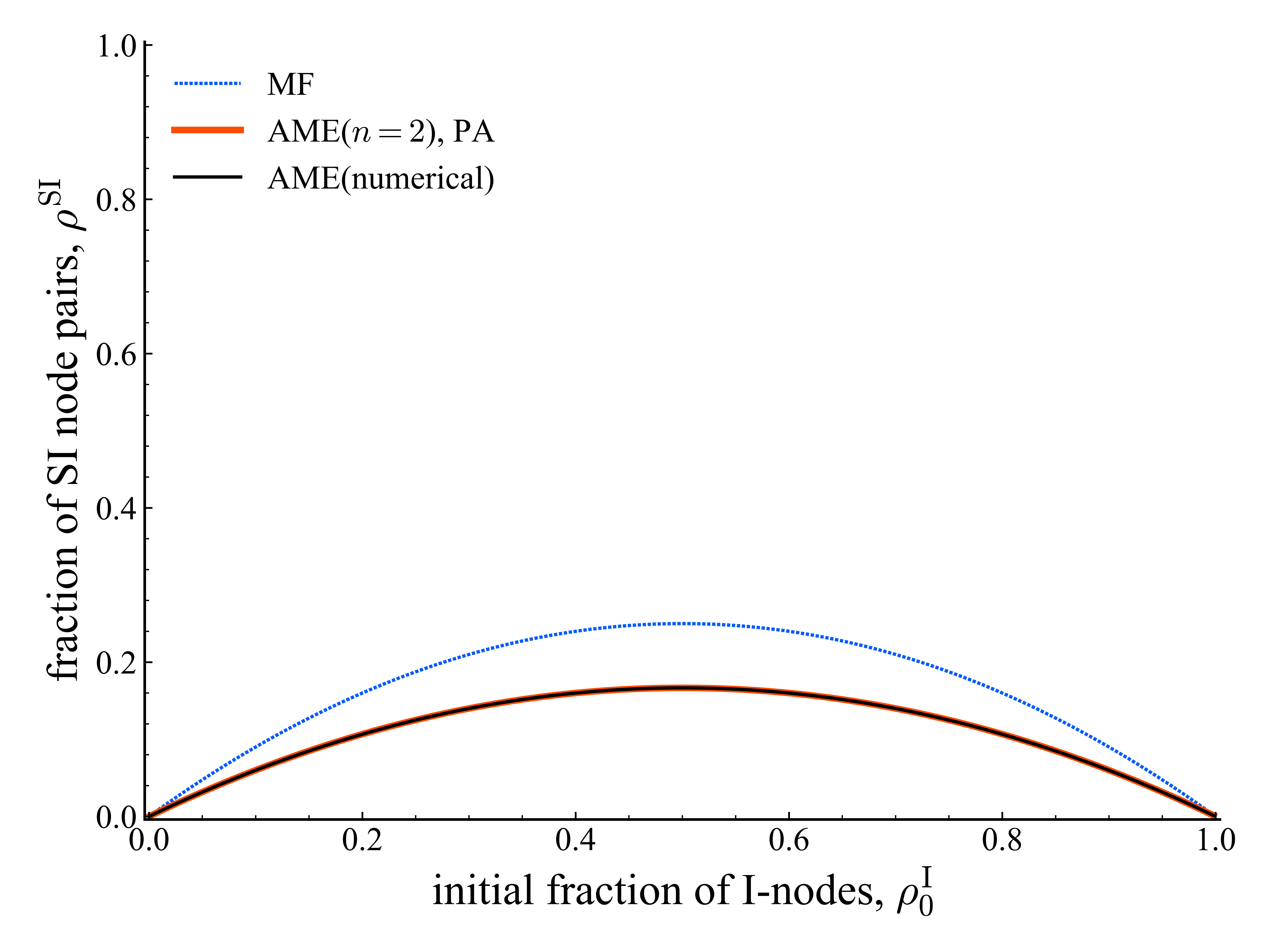}
    \caption{
        Steady state of the voter model with degree $k=4$.
        MF and PA denote the mean-field and pair approximations, respectively.
        AME $(n=2)$ denote the steady state of the AME with moments up to the second order.
	    The black line represents the numerical solution of the AME.
        \label{fig:voter_model}
    }
\end{figure}

We consider the voter model, a simple model for studying consensus formation. 
Each node can be in one of two states, denoted by $\S$ and $\I$, representing the two possible opinions.
State update events occur at a constant rate $1$, during which a randomly selected node imitates the state of one of its neighbors.
The transition rates are given by $W(\S_m \to \I_m) = m/k$ and $W(\I_m \to \S_m) = (k-m)/k$. 
In this model, the time evolution of $\rho^{\I}$ is given by $\dv{t} \rho^{\I} = \ev*{m}_{\S} - \ev*{k-m}_{\I}$.
Due to the symmetry between SI and IS node pairs, $\ev*{m}_{\S} = \ev*{k-m}_{\I}$, the fraction of $\I$-nodes remains constant over time: $\rho^{\I}(t) = \rho^{\I}_0$.
While the voter model on a finite system eventually reaches consensus (i.e., $\rho^{\I} = 0$ or $1$), the AME describes the dynamics in the infinite-size limit, where fluctuations vanish and the system instead remains in a stationary state without reaching consensus.

From Eqs.~\eqref{eq:moment_jS} and~\eqref{eq:moment_jI} for the time evolution of the moments of order $j=0, 1, 2$, 
the following five independent equations hold in the steady state:
\begin{align}
    1 &= \ev*{m^0}_{\S} + \ev*{m^0}_{\I},
    \label{eq:moment_voter_1}\\
    0 &= \ev*{m}_{\S} - \ev*{k-m}_{\I},
    \label{eq:moment_voter_2}\\
    0 &= \ev*{m^2}_{\S} - \ev*{m(k-m)}_{\I},
    \label{eq:moment_voter_3}\\
    0 &= 
    - \ev*{ m^3 }_{\S} + \ev*{ m^2 (k-m) }_{\I}, \nonumber\\
    &\qquad
    + 2 \frac{ \qty( \ev*{ m (k-m) }_{\S} )^2 }{ \ev*{ k-m }_{\S} } + \ev*{ m (k-m) }_{\S} 
    - 2 \frac{ \ev*{ (k-m)^2 }_{\I} \ev*{ m^2 }_{\S} }{ \ev*{ k-m }_{\I} } + \ev*{ (k-m)^2 }_{\I},
    \label{eq:moment_voter_4}\\
    0 &=
    - \ev*{ m^2 (k-m) }_{\I} + \ev*{ m^3 }_{\S} \nonumber\\
    &\qquad
    + 2 \frac{ \ev*{ m^2 }_{\S} \ev*{ m (k-m) }_{\I}}{ \ev*{ m }_{\S} } + \ev*{ m^2 }_{\S} 
    - 2 \frac{ \ev*{ m (k-m) }_{\I} \ev*{ m^2 }_{\I} }{ \ev*{ m }_{\I} } + \ev*{ m (k-m) }_{\I}.
    \label{eq:moment_voter_5}
\end{align}
Note that Eq.~\eqref{eq:moment_voter_2} is doubly degenerate.
The above set of equations is not closed because it involves third-order moments. 
Using Eq.~\eqref{eq:approximation} with $n=2$,
\begin{align}
    0 &= \ev*{ m^3 }_{\S} - \ev*{ m^2 (k-m) }_{\I},
    \label{eq:moment_voter_6}
\end{align}
we obtain the stationary fraction of SI node pairs, $\rho^{\S\I} = \ev*{m}_{\S}/k$, as
\begin{align}
\rho^{\S\I} &= \frac{k-2}{k-1} \rho^{\I}_0 \qty( 1 - \rho^{\I}_0 ),
\end{align}
which coincides with that obtained via the pair approximation.
Figure \ref{fig:voter_model} shows the steady states obtained by each approximation.
For the voter model, the $n=2$ result (red solid line) is nearly identical to the exact solution of the AME (black solid line). 
This suggests that the steady state of the voter model satisfies the semi-detailed balance condition.

\subsection{Evolutionary Game}

\begin{figure}[htbp]
    \centering
    \includegraphics[width=0.8\linewidth]{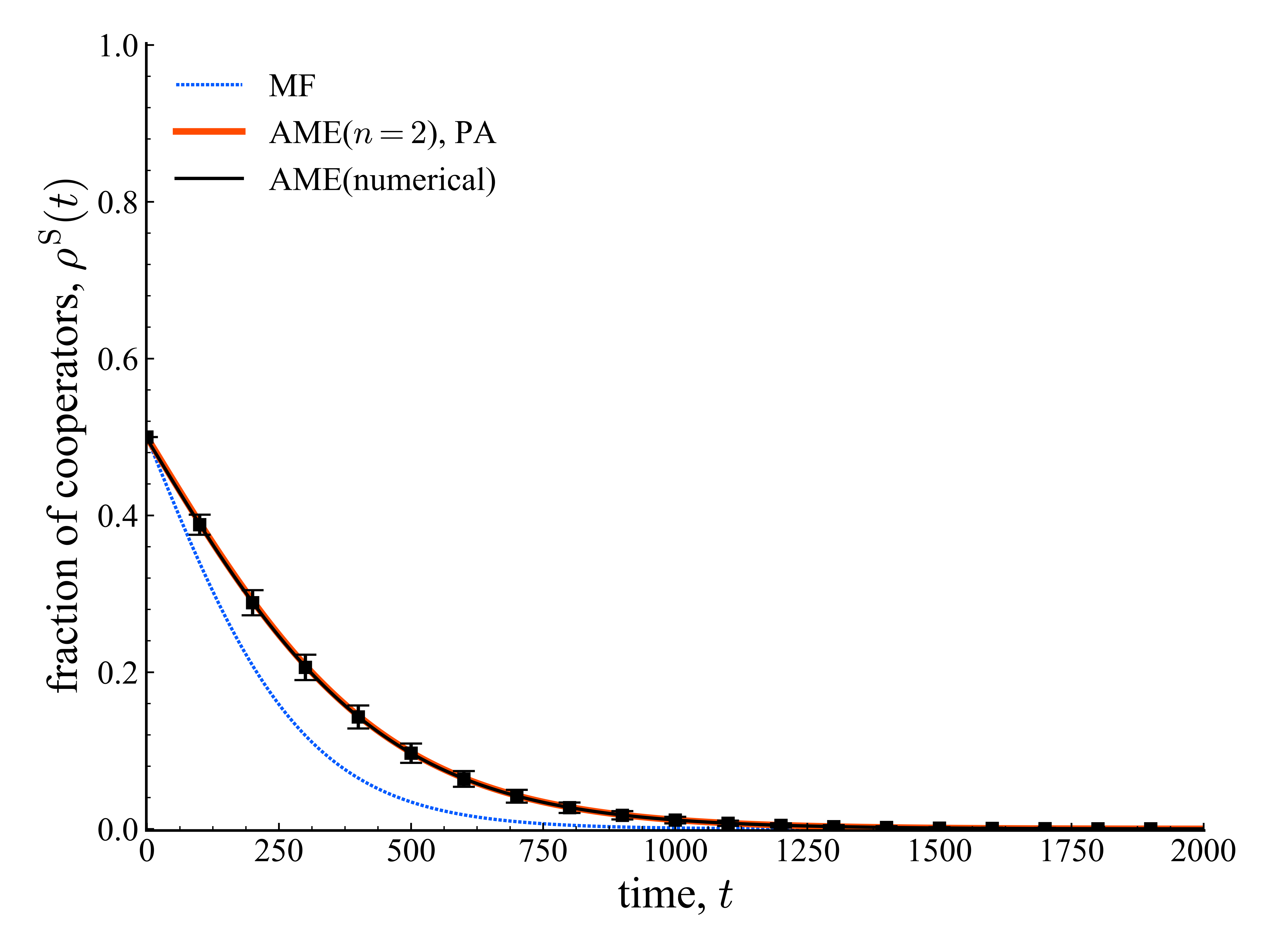}
    \caption{
        Time evolution of the Prisoner's Dilemma game
        with degree $k=4$, benefit $b=2$, cost $c=1$, and intensity of selection $\beta=1/300$.
        MF and PA denote the mean-field and pair approximations, respectively.
        AME $(n=2)$ represents the steady state of the AME approximated by considering moments up to the second order.
        The black solid line represents the numerical solution of the AME.
        Black squares represent Monte Carlo simulation results on a regular random graph with $10^5$ nodes. 
        Monte Carlo results are averaged over $10^3$ independent runs, with error bars indicating the standard deviation.
        \label{fig:game}
    }
\end{figure}

We consider a $2 \times 2$ game on a network, where each node represents a player and edges connect interacting individuals. 
Each player adopts one of two strategies, $\S$ or $\I$.
Let $\pi_{ss'}$ denote the payoff of an $s$-node interacting with an $s'$-node. 
The total payoff of an $s_m$-node is given by $\Pi_{s_m} \equiv m \pi_{s \I} + (k-m) \pi_{s \S}$. 
We employ the pairwise comparison rule~\cite{Szabo2005,Ohtsuki2006,Perc2017,Takiguchi2024}, a standard strategy update rule. 
In this rule, a focal node is randomly selected to update its strategy with a constant rate $1$.
This node selects a neighbor at random as a role model and imitates its strategy with a probability determined by their payoffs.
When the focal node is in state $s_m$ and the role model is in state $s'_{m'}$, the imitation probability is given by a sigmoid function: $\qty(1+\exp \qty[\beta \qty(\Pi_{s_m} - \Pi_{s'_{m'}})])^{-1}$, where $\beta$ controls the intensity of selection. 
The transition rates are given by~\cite{Takiguchi2024}:
\begin{align}
    W(\S_m \to \I_m) &= \frac{m}{k} \sum_{m'=0}^k \frac{(k-m') \rho^{\I}_{m'}}{\sum_{m''=0}^k (k-m'') \rho^{\I}_{m''}} \frac{1}{1 + \e^{\beta \qty(\Pi_{\S_m} - \Pi_{\I_{m'}})}},\\
    W(\I_m \to \S_m) &= \frac{k-m}{k} \sum_{m'=0}^k \frac{m' \rho^{\S}_{m'}}{\sum_{m''=0}^k m'' \rho^{\S}_{m''}} \frac{1}{1 + \e^{\beta \qty(\Pi_{\I_m}- \Pi_{\S_{m'}})}}.
\end{align}
Here, $(k-m') \rho^{\I}_{m'} / \sum_{m''=0}^k (k-m'') \rho^{\I}_{m''}$ represents the probability that an $\I$-node (the role model) adjacent to an $\S$-node (the focal node) is in state $\I_{m'}$, and $m' \rho^{\S}_{m'} / \sum_{m''=0}^k m'' \rho^{\S}_{m''}$ represents the probability that an $\S$-node adjacent to an $\I$-node is in state $\S_{m'}$.

Note that the models considered so far have transition rates linear in $m$, whereas the present model has nonlinear transition rates.
Using the singular perturbation method~\cite{Ohtsuki2006,Takiguchi2024,Takiguchi2026}, we show that the dynamics become analytically tractable in the weak-selection limit ($\beta \ll 1$).
By taking the Taylor expansion of the sigmoid function with respect to $\beta$, the transition rates are given by
\begin{align}
    W(\S_m \to \I_m) &= \frac{1}{2} \frac{m}{k} + \frac{\beta}{4} \frac{m}{k} \sum_{m'=0}^k \frac{(k-m') \rho^{\I}_{m'}}{\sum_{m''=0}^k (k-m'') \rho^{\I}_{m''}} \qty(\Pi_{\I_{m'}} - \Pi_{\S_m}) + \order{\beta^2},\\
    W(\I_m \to \S_m) &= \frac{1}{2} \frac{k-m}{k} + \frac{\beta}{4} \frac{k-m}{k} \sum_{m'=0}^k \frac{m' \rho^{\S}_{m'}}{\sum_{m''=0}^k m'' \rho^{\S}_{m''}} \qty(\Pi_{\S_{m'}} - \Pi_{\I_m}) + \order{\beta^2}.
\end{align}
The zeroth-order terms in $\beta$ correspond to the transition rates of the voter model. 
The factor of $1/2$ merely affects the relaxation speed and can thus be neglected when considering the steady state. 
The time evolution of $\rho^s_m$ is of order $\beta^0$, indicating that these variables rapidly relax to the steady state of the voter model. 
In contrast, the time evolution of the fraction of $\S$-nodes, $\rho^{\S}$, occurs on a slower timescale of order $\beta^1$:
\begin{align}
    \dv{t} \rho^{\S} 
    = 
    &- \frac{\beta}{4} \sum_{m=0}^k \frac{m}{k} \rho^{\S}_m \sum_{m'=0}^k \frac{(k-m') \rho^{\I}_{m'}}{\sum_{m''=0}^k (k-m'') \rho^{\I}_{m''}} \qty(\Pi_{\I_{m'}} - \Pi_{\S_m}) \nonumber\\
    &+ \frac{\beta}{4} \sum_{m=0}^k \frac{k-m}{k} \rho^{\I}_m \sum_{m'=0}^k \frac{m' \rho^{\S}_{m'}}{\sum_{m''=0}^k m'' \rho^{\S}_{m''}} \qty(\Pi_{\S_{m'}} - \Pi_{\I_m})  + \order{\beta^2}.
\end{align}
By exploiting this separation of timescales and substituting the steady-state relations~\eqref{eq:moment_voter_1}--\eqref{eq:moment_voter_3} of the voter model, we obtain:
\begin{align}
    \dv{t} \rho^{\S} 
    = 
    &- \frac{\beta}{4k} \sum_{m'=0}^k (k-m') \rho^{\I}_{m'} \Pi_{\I_{m'}}
    + \frac{\beta}{4k} \sum_{m=0}^k m \rho^{\S}_m \Pi_{\S_m} \nonumber\\
    &+ \frac{\beta}{4k} \sum_{m'=0}^k m' \rho^{\S}_{m'} \Pi_{\S_{m'}}
    - \frac{\beta}{4k} \sum_{m=0}^k (k-m) \rho^{\I}_m \Pi_{\I_m} \\
    = 
    &+ \frac{\beta}{2} \ev*{m}_{\S} (\pi_{\S\S} - \pi_{\I\S}) 
    + \frac{\beta}{2k} \ev*{m^2}_{\S} \qty(\pi_{\I\S} - \pi_{\S\S} + \pi_{\S\I} - \pi_{\I\I}  ).
\end{align}
Note that approximation~\eqref{eq:moment_voter_6} has not been used in this derivation.

As an example, we consider the Prisoner's Dilemma game, in which strategies $\S$ and $\I$ denote cooperation and defection, respectively.
The payoff matrix is given by
\begin{align}
    \begin{pmatrix}
        \pi_{\S\S} & \pi_{\S\I}\\
        \pi_{\I\S} & \pi_{\I\I}
    \end{pmatrix}
    =
    \begin{pmatrix}
        b-c & -c\\
        b & 0
    \end{pmatrix},
\end{align}
where $c$ ($>0$) represents the cost of cooperation and $b$ ($>c$) is the benefit provided by a cooperator. 
Substituting these payoffs into the above expression, we obtain the time evolution of $\rho^{\S}$ as
\begin{align}
    \dv{t} \rho^{\S} = - \frac{\beta}{2} \ev*{m}_{\S} c < 0,
\end{align}
implying that cooperators become extinct in the steady state, $\rho^{\S}=0$.
This result is consistent with that obtained by Ohtsuki and Nowak~\cite{Ohtsuki2006} using the pair approximation. 
Our framework shows that this result remains valid even when higher-order correlations are incorporated through the AME.

Using Eqs.~\eqref{eq:moment_voter_1}--\eqref{eq:moment_voter_5} together with the $n=2$ approximation~\eqref{eq:moment_voter_6}, we obtain the following time evolution:
\begin{align}
    \dv{t} \rho^{\S} 
    &= 
    \beta \frac{(k-2)^2}{2(k-1)} \rho^{\S} (1-\rho^{\S})
    \qty[
        \qty(\pi_{\S\S} - \pi_{\I\S} + \pi_{\I\I} - \pi_{\S\I}) \rho^{\S}
        + \frac{\pi_{\I\S} - \pi_{\S\S} + (k-1) \qty(\pi_{\S\I} - \pi_{\I\I})}{k-2}
    ]\\
    &= 
    \beta \frac{(k-2)^2}{2(k-1)} \rho^{\S} 
    \qty[
        \sum_{s'\in\{\S,\I\}} \rho^{s'} \pi_{\S s'}
        + \sum_{s'\in\{\S,\I\}} \rho^{s'} \frac{\pi_{\S\S} + \pi_{\S s'} - \pi_{s'\S} - \pi_{s's'}}{k-2}
        - \sum_{s,s'\in\{\S,\I\}} \rho^{s} \rho^{s'} \pi_{ss'}
    ].
\end{align}
This expression is identical to that obtained via the pair approximation~\cite{Ohtsuki2006}. 
For example, in the Prisoner's Dilemma game, the evolution equation reduces to
\begin{align}
    \dv{\rho^{\S}}{t} 
    &= - \beta \frac{k(k-2)}{2(k-1)} c \rho^{\S} \qty( 1 - \rho^{\S} ).
\end{align}
This equation can be solved analytically as
\begin{align}
    \rho^{\S} (t) &= \frac{1}{1 + (1/\rho^{\S}_{0}-1) \e^{\beta \frac{k(k-2)}{2(k-1)} c t}}.
\end{align}
Figure~\ref{fig:game} shows the time evolution of the Prisoner's Dilemma game under weak selection. 
The mean-field approximation (blue dotted line) exhibits a significant deviation from the numerical solution of the AME (black solid line).
In contrast, the $n=2$ result (red solid line) is in excellent agreement with both the numerical solution of the AME and the Monte Carlo simulation results (black squares).

\section{Discussion and Conclusion}

We have developed an analytical framework to determine the steady states of dynamical processes on networks with an accuracy at least as high as that of the pair approximation. 
By systematically refining the approximation, the steady states obtained by our framework approach the exact solutions of the AME.
We have uncovered previously unrecognized aspects of both the pair approximation and the AME through the introduction of the ``semi-detailed balance condition''.
The pair approximation satisfies this condition, under which the steady state is fully characterized by the mean and variance of $m$, the number of neighbors in state $\I$. 
In contrast, the AME achieves higher accuracy by incorporating higher-order moments.
We expect that our framework will contribute to a deeper mathematical understanding of dynamical processes on networks and their approximation methods.

Our framework is applicable to models in which the transition rates are linear in $m$, the number of $\I$-neighbors. 
This class includes, in addition to the SIS and voter models, the Bass diffusion model, the Kirman model, and Ising Glauber dynamics at high temperatures~\cite{Gleeson2013}. 
The framework can also be extended to models with transition rates given by finite-order polynomials of $m$. 
In contrast, it is not applicable to models with discontinuous transition rates, such as the Watts threshold model and Ising Glauber dynamics at the zero-temperature limit.

The framework can be extended to models with three or more node states such as the SIR model.
It can also be extended to networks with arbitrary degree distributions. 
However, in these generalizations, the number of distinct moments increases combinatorially with the order, making it difficult to obtain the steady state.

For evolutionary games, we have derived analytical solutions for both the steady states and the time evolution by combining our proposed framework with the singular perturbation method. 
Our results confirm that the conditions for the evolution of cooperation, originally derived under the pair approximation, remain robust even under the higher-order accuracy of the AME.
Although this study has focused on the pairwise comparison rule, the same approach can be applied to other strategy update rules, such as birth-death or death-birth rules.

\section*{Acknowledgment}

This work was supported by JST SPRING (JPMJSP2119), JSPS KAKENHI (24K06879) and JST ERATO SAKAI Real and Abstract Gels Project (JPMJER2401).


\appendix

\section{Moment Relations under the Pair Approximation}
\label{sec:pair_approximation}

In this appendix, we show that under the pair approximation, the following relation holds for any $j=0,1,\dots$ in the steady state:
\begin{align}
    - \ev*{ m^j W(\I_m \to \S_m) }_{\I} + \ev*{ m^j W(\S_m \to \I_m) }_{\S} = 0
    \label{eq:approximation_pair_appendix}.
\end{align}
From Eq.~\eqref{eq:moment_jI}, we obtain the following steady-state relations for $j=0$ and $j=1$:
\begin{align}
    - \ev*{ W(\I_m \to \S_m) }_{\I} + \ev*{ W(\S_m \to \I_m) }_{\S} &= 0,\\
    - \ev*{ m W(\I_m \to \S_m) }_{\I} + \ev*{ m W(\S_m \to \I_m) }_{\S} &= 0.
    \label{eq:moment_1I}
\end{align}
These relations hold not only under the pair approximation but also in the AME.
For $j \geq 2$, by substituting Eq.~\eqref{eq:moment_1I} into Eq.~\eqref{eq:moment_jI}, we obtain the following steady-state relation:
\begin{align}
    0 =
    &- \ev*{ m^j W(\I_m \to \S_m) }_{\I} + \ev*{ m^j W(\S_m \to \I_m) }_{\S}\nonumber\\
    &+ \ev*{ m W(\I_m \to \S_m) }_{\I} \ev*{k-m}_{\I} \ev*{m}_{\I} \nonumber\\
    &\qquad \times \qty[ \ev*{m}_{\I} \ev*{ \qty[ (m+1)^j - m^j ] (k-m) }_{\I} - \ev*{k-m}_{\I} \ev*{ \qty[ m^j - (m-1)^j ] m }_{\I} ].
\end{align}
Therefore, it suffices to show that
\begin{align}
    X \equiv \ev*{m}_{\I} \ev*{ \qty[ (m+1)^j - m^j ] (k-m) }_{\I} - \ev*{k-m}_{\I} \ev*{ \qty[ m^j - (m-1)^j ] m }_{\I} = 0.
    \label{eq:balance_condition_pair}
\end{align}
Here, $X$ is expressed as
\begin{align}
    X =
    &  \sum_{m_1 =0}^k m_1 \rho^{\I}_{m_1} \sum_{m_2 =0}^k \qty[ (m_2 +1)^j - m_2^j ] (k-m_2) \rho^{\I}_{m_2} \nonumber\\
    & - \sum_{m_2=0}^k (k-m_2) \rho^{\I}_{m_2} \sum_{m_1 =0}^k \qty[ m_1^j - (m_1-1)^j ] m_1 \rho^{\I}_{m_1}.
\end{align}
Recalling that under the pair approximation, $\rho^{\I}_m = \rho^{\I} B_m(\rho^{\I|\I})$, where $B_m(\cdot)$ denotes the binomial distribution~\eqref{eq:binomial}, we obtain:
\begin{align}
    X =
    & \qty(\rho^{\I})^2 \sum_{m_1 =0}^k \sum_{m_2 =0}^{k} \frac{k!}{m_1! (k-m_1)!} \frac{k!}{m_2! (k-m_2)!} \qty(\rho^{\I|\I})^{m_1+m_2} \qty(1-\rho^{\I|\I})^{2k-m_1-m_2} \nonumber\\
    &\quad \times m_1 (k-m_2) \qty[ (m_2 +1)^j + (m_1-1)^j - m_1^j - m_2^j ]\\
    =
    & \qty(\rho^{\I})^2 \sum_{m_1 =1}^k \sum_{m_2 =0}^{k-1} \frac{k!}{(m_1-1)! (k-m_1)!} \frac{k!}{m_2! (k-m_2-1)!} \qty(\rho^{\I|\I})^{m_1+m_2} \qty(1-\rho^{\I|\I})^{2k-m_1-m_2}\nonumber\\
    &\quad \times \qty[ (m_2 +1)^j + (m_1-1)^j - m_1^j - m_2^j ].
\end{align}
By changing dummy index $m_1' = m_1 - 1$ and $m_2' = m_2 + 1$, we obtain
\begin{align}
    X=&
    \qty(\rho^{\I})^2 
    \sum_{m_1' =0}^{k-1} \sum_{m_2'=1}^{k} \frac{k!}{m_1'! (k-m_1'-1)!} \frac{k!}{(m_2'-1)! (k-m_2')!} \qty(\rho^{\I|\I})^{m_1'+m_2'} \qty(1-\rho^{\I|\I})^{2k-m_1'-m_2'}\nonumber\\
    &\quad \times \qty[ m_2'^j + m_1'^j - (m_1'+1)^j - (m_2'-1)^j ] = - X.
\end{align}
This implies that $X=0$.

\section{Equivalence Between the Moment Relation and the Semi-detailed Balance Condition}
\label{sec:balance_condition}

In this appendix, we show that the following moment relation:
\begin{align}
    - \ev*{ m^j W(\S_m \to \I_m) }_{\S} + \ev*{ m^j W(\I_m \to \S_m) }_{\I} = 0
\end{align}
for any $j = 0, 1, \ldots$, is equivalent to the semi-detailed balance condition:
\begin{align}
    - W(\S_m \to \I_m) \rho^{\S}_m + W(\I_m \to \S_m) \rho^{\I}_m = 0
\end{align}
for any $m = 0, 1, \ldots, k$.

Recalling that $\ev*{f(m)}_s = \sum_m f(m)\rho_m^s$, the latter immediately implies the former by multiplying both sides by $m^j$ and summing over $m$.
To prove the converse, we introduce the following generating functions:
\begin{align}
    M_{\S}(y) &\equiv \sum_{m=0}^k \e^{ y m } W(\S_m \to \I_m) \rho^{\S}_m,\\
    M_{\I}(y) &\equiv \sum_{m=0}^k \e^{ y m } W(\I_m \to \S_m) \rho^{\I}_m.
\end{align}
Expanding $\e^{ym}$ in a Taylor series, we obtain
\begin{align}
    M_{\S}(y) 
    &= \sum_{m=0}^k \qty( \sum_{j =0}^\infty \frac{y^j m^j}{j!} ) W(\S_m \to \I_m) \rho^{\S}_m \\
    &= \sum_{j =0}^\infty \frac{y^j }{j!} \ev*{ m^j W(\S_m \to \I_m) }_{\S}\\
    &= \sum_{j =0}^\infty \frac{y^j }{j!} \ev*{ m^j W(\I_m \to \S_m) }_{\I}
    = M_{\I}(y),
\end{align}
where we used the moment relation.
Since these are finite sums, equality of the generating functions implies equality of the corresponding coefficients for all $m$.
Therefore, we conclude that $W(\S_m \to \I_m) \rho^{\S}_m = W(\I_m \to \S_m) \rho^{\I}_m$ holds for all $m$.


\bibliographystyle{unsrt}
\bibliography{bib}

\end{document}